\documentclass[a4paper,12pt]{article}
\usepackage{graphicx,calc}
\usepackage{minitoc}
\usepackage[english]{babel}
\usepackage{float}
\usepackage{amsmath}
\usepackage{tikz}
\usepackage[utf8]{inputenc} 
\usepackage{multicol}
\renewcommand{\theequation}{\arabic{section}.\arabic{equation}}

\newcommand{\be}{\begin{equation}}
\newcommand{\ee}{\end{equation}}
\newcommand{\ba}{\begin{array}}
\newcommand{\ea}{\end{array}}
\newcommand{\bc}{\begin{center}}
\newcommand{\ec}{\end{center}}
\newcommand{\disregard}[1]{{}}

\newcommand{\ds}{\displaystyle}

\newcommand{\demi}{{\ds 1\over\ds 2}}

\begin{document}

\title{Binary Mixture of quasi one dimensional dipolar Bose Einstein Condensates with tilted dipoles}
\author{ Ahmed  Hocine and Mohamed Benarous \\
{\it Laboratory for Theoretical Physics and Material Physics} \\
{\it Faculty of Exact and Computer Sciences} \\
{\it Hassiba Benbouali University of Chlef (Algeria)} \\}

\date{\today}
\maketitle

\begin{abstract}
We consider a $^{168}$Er-$^{164}$Dy dipolar mixture, trapped by a cigar shaped harmonic potential. 
We derive the quasi-1D inter-species effective potential exhibiting the tilting angles and show that 
it is a quite natural generalization of the situation of a single dipolar gas. By solving the coupled 
Gross-Pitaevskii equations, we observe a transition from miscible to immiscible mixture as the orientations 
of the magnetic moments are varied. The atom numbers are also shown to lead to noticeable effects on the mixture. 
\end{abstract}

\section{Introduction}

The nature of inter particle interactions in ultracold gases at low-dimensions has been extensively studied 
theoretically and experimentally over the last years\cite{Petrov, Santos}. In one and two dimensions, physics is qualitatively (and quantitatively) different and new phenomena are observed \cite{BKT, Paredes, baranov, Lahaye}. 

The properties of quantum gases are crucially determined by the leading role of interactions, which are in turn 
affected by a constrained geometry. This can be bestly seen when the interactions are anistropic and/or long ranged\cite{Deuretzbacher, yukalov, mulkerin, baillie, Pfau}. For instance, for dipolar gases, not only the amplitude 
but also the sign of the interactions are strongly affected by the trapping geometry. Indeed, the stability diagram 
contains the trap anisotropy as a crucial parameter\cite{Lahaye, yiyou, dutta, santos2, ronen, edler, martinblakie}.

Binary mixtures of cold atoms or cold molecules show a richer phase diagram than a single species gas owing to the 
existence of many different interactions. Indeed, in addition to the intra-species interactions, the inter-species 
forces have been predicted and then shown experimentally to play a prominent role in the dynamics of the mixture\cite{mineevall, chw}. 

A relevant characteristic of binary mixtures is their ability to mix and demix depending on various parameters, 
such as the intra and inter-species interactions, the number of particles as well as the trapping geometries.
In this context, much have been done in the full three dimensional as well as in the quasi-2D and quasi-1D cases 
with dipolar gases\cite{adhikari2, Adhikari, kumar, Adhikari2014} with the predictions of structure formation due essentially to the dipolar interactions. This mixing-demixing behavior is bestly described by the overlap of the 
macroscopic wave functions and has recently attracted attention\cite{kumar, wen}.

For dipolar binary mixtures, another tuning parameter that governs the miscibility-immiscibility 
transition as well as the stability regions is the orientation of the magnetic moment. Furthermore, 
it may also be used to control the anisotropy by leading to a rotonlike Bogoliubov spectrum \cite{baillie} 
and may drive the condensate to a phonon instability\cite{santos3}. 

Except from few works on single dipolar gases, such as \cite{santos3} who have considered 2D bright 
solitons in a dipolar BEC, or \cite{mishra} who discussed a quasi-2D BEC with tilted dipoles, the studies 
of dipolar bose mixtures under various orientations of the magnetic moments for strong anisotropies are, 
to our knowledge, just at their beginning. As a first example, we can cite in particular the work of 
\cite{gligoric} where one of the key results is that "the long-range repulsion tends to suppress the 
spatial structure induced by the immiscibility, while the nonlocal attraction helps to enhance it". 
We will show in our numerical treatment that it is indeed the case, namely that changing the orientations 
of the dipoles and therefore changing the interactions from attractive to repulsive, induces a clear tendency 
to mixing. 

Moreover, when studying a $^{168}$Er-$^{164}$Dy mixture, the authors of \cite{kumar} have found that 
the mixture is {\it partially} miscible (as compared to for instance a $^{164}$Dy-$^{162}$Dy mixture 
which is {\it completely} miscible), and this was attributed to the inter-species dipolar strength. 
The stability window, in terms of the fraction of atom numbers, was also found to be very narrow 
for pancake traps. In our case, we will show that, not only the stability window can be wider for 
cigare-shaped traps, accomodating higher numbers of atoms in each species and therefore almost reaching 
the Thomas-Fermi regime, but also that the mixture can be made completely miscible by just changing 
the orientations of the magnetic moments of the two species while keeping fixed all the other parameters. 
We will alo demonstrate, based on energy arguments, that it is the less energetic species, with an attractive 
dipolar component, that will occupy the center of the trap. By contrast, it is noted in \cite{Adhikari2014}
that "the species containing the larger number of atoms stay at the center and the species containing the smaller 
number of atoms breaks into two equal parts and stays symmetrically on two sides with a minimum of interspecies
overlap". This claim is however parameter and model dependent. Indeed, in the previous reference, the author 
considers first that the two species have different scattering lengths and then gradually lowers the axial trap to 
get an axially free gas.

In order to extend the discussions started in the preceding quoted papers, we aim in this work at examining more 
deeply how a dipolar binary mixture is affected by the changes of the orientations of the magnetic moments. More specifically, owing to the actual experimental studies\cite{aikawa, lu} and to the high magnetic moments of the atoms involved, we choose a mixture composed of $^{168}$Er and $^{164}$Dy gases.

We will focus on a quasi-1D geometry since it provides a strong anisotropic problem which will help us to 
apprehend better the role of the tilting angles. Indeed, varying the orientations of the magnetic moments 
may affect not only the sign of the dipole-dipole interactions (DDI), changing them from repulsive to 
attractive, but also the anisotropy of the system even if everything else is kept fixed. This was shown by 
Gligoric et al.\cite{gligoric} in the two extreme limits (side by side and head to tail). Furthermore, the 
DDI may drastically change the properties of the system in quasi-1D even if the short range interactions 
dominate the physics in the full 3D geometry. 

To this end, we construct first the quasi-1D effective potential for both the intra-species and the 
inter-species interactions. While the former has been derived elsewhere, we show that the latter  
nicely generalizes the expression derived in \cite{Santos} to the case of mixture of different species 
and depending on both orientations of the magnetic moments, expression which was not {\it a priori} 
foreseeable since it depends on both the relative angle and the total angle. 

The paper is organized as follows. In the next section, we present the zero temperature model for a trapped binary mixture of dipolar gases in the mean field approach and obtain its quasi-one dimensional reduction when the transverse degrees of freedom are frozen.

Section 3 is devoted to the numerical resolution of the two coupled Gross-Pitaevskii equations. 
The long range character of the DDI makes the numerical problem (being an integro-differenial one) quite 
challenging and we build a special algorithm for it. We then study the behavior of the densities of each 
species in terms of the orientations of the magnetic moments. We also examine their individual energies and 
show that the species with lower energy and an attractive dipolar component does indeed occupy the center of 
the trap. We then observe how the mixing and demixing occurs as one varies the tilting angles. 
An interesting parameter is a measure of the overlap between the wave functions which we define and plot as a 
function of the tlting. As it can easily be understood, this parameter is seen to be minimum for a head-to-tail 
configuration while it is maximum for a side by side configuration. Energy arguments provide simple explanations of 
these behaviors. But since energy depends  also on the number of atoms in each species, it is quite natural to pursue 
an analogous analysis along the same preceding lines by examining the role of the number of atoms in the mixing-demixing process.

The last section summarises our conclusions and provides some perspectives for forthcoming works.

\section{The model}

In the following, we consider a dipolar Bose gas composed of two-species, with inter and intra-species dipolar 
and contact interactions. The system is enclosed in a cigar-shaped harmonic potential, which is an axially-symetric 
trap with strong transverse confinement, generated by a deep optical lattice. The confinement is ensured by two 
harmonic potentials $V_{ex_{i}}=\frac{m_{i}}{2}(\omega^{2}_{i}x^{2}+{\omega}_{i\bot}^2\rho^{2})$ ($i=1,2$), $\omega_{i}$ 
and $\omega_{i\bot}$ being the trap frequencies of the axial ($\vec{x}$) and transverse ($\vec{\rho}$) directions respectively, with $\omega_{i\bot}>>\omega_{i}$ in order to freeze the transverse degrees of freedom. 
In the mean-field approximation, the hamiltonian of the system writes 
\be
\ba{rl}
 H&=\sum_{i=1}^2
 \int d\textbf{r}\Psi_i^{+}(\textbf{r},t)(-\frac{\hbar^{2}}{2m_{i}}\nabla^2+V_{ex_{i}}(\textbf{r}))\Psi_i(\textbf{r},t) \\
 &+\sum_{i=1}^2
 \int d\textbf{r} d\textbf{r}'\Psi_i^{+}(\textbf{r},t)\Psi_i^{+}(\textbf{r}',t)U_{i}(\textbf{r},\textbf{r}')\Psi_i(\textbf{r}',t)\Psi_i(\textbf{r},t)
 \\
  &+\int d\textbf{r} d\textbf{r}'\Psi_1^{+}(\textbf{r},t)\Psi_2^{+}(\textbf{r}',t)U_{21}(\textbf{r},\textbf{r}')\Psi_2(\textbf{r}',t)\Psi_1(\textbf{r},t)
\\  
  &+\int d\textbf{r} d\textbf{r}'\Psi_2^{+}(\textbf{r},t)\Psi_1^{+}(\textbf{r}',t)U_{12}(\textbf{r},\textbf{r}')\Psi_1(\textbf{r}',t)\Psi_2(\textbf{r},t)\\

\ea
\label{eq1}
\ee
where $\Psi_i^{+}(\textbf{r},t)$ and $\Psi_i(\textbf{r},t)$ are the boson creation and anihilation operators for the species $i$, $U_{i}(\textbf{r},\textbf{r}')$ are the intra-species potentials 
\be
U_{i}(\textbf{r},\textbf{r}')=\frac{4\pi\hbar^{2}a_{i}}{m_{i}}\delta(\textbf{r}-\textbf{r}')+\frac{\mu_{0} \mu_{i }^{2}}{4\pi}U_{dd_{i}}^{(3D)}(\textbf{r}-\textbf{r}'),
\label{eq211}
\ee
and $U_{12}(\textbf{r},\textbf{r}')=U_{21}(\textbf{r}',\textbf{r})$ the inter-species potential 
\be
U_{12}(\textbf{r},\textbf{r}')=\frac{2\pi\hbar^2 a_{12}}{m_r}\delta(\textbf{r}-\textbf{r}')+\frac{\mu_0\mu_1\mu_2}{4\pi}U_{dd_{12}}^{(3D)}(\textbf{r}-\textbf{r}'), 
\label{eq2}
\ee
where we have considered the most general contact+dipolar interactions. In the previous expressions, $m_i$, $a_i$ and $\mu_i$ are the masses, the s-wave scattering lengths and the magnetic moments of the two species, $m_r=\frac{m_1 m_2}{m_1+m_2}$ the reduced mass, $a_{12}$ the s-wave scattering length corresponding to the binary interaction. $\mu_0$ is the permeability of free space. 

The dipolar potentials $U_{dd_{i}}^{(3D)}$ and $U_{dd_{12}}^{(3D)}$ are defined as usual by 
\be
\ba{rl}
U_{dd_{i}}^{(3D)}(\textbf{r}-\textbf{r}')&=\frac{1-3\cos^2\theta_{i}}{\lvert \textbf{r}-\textbf{r}'\lvert ^{3}}, \\
U_{dd_{12}}^{(3D)}(\textbf{r}-\textbf{r}')&=\frac{\cos(\theta_{1}-\theta_{2})-3\cos\theta_1\cos\theta_2}{\lvert \textbf{r}-\textbf{r}'\lvert ^{3}},
\ea
\label{eq331}
\ee
where $\theta_i$ are the angles formed by the vector $\textbf{r}-\textbf{r}'$ and the magnetic moments $\vec{\mu}_i$ (see appendix).

Upon introducing the aspect ratios $\lambda_{i}=\frac{\omega_{i}}{\omega_{i\bot}}$ and using the notations of \cite{Adhikari}, the coupled Gross-Pitaevskii equations (GPE) write as 
\be
\ba{rl}
i \frac{\partial \Psi_1(\textbf{r},t)}{\partial t}=&
\big\{
-\frac{1}{2}\nabla^2+\frac{1}{2}(\lambda_1^2 x^2+\rho^2)+g_1^{(3D)}\lvert\Psi_1(\textbf{r},t)|^{2}+g_{dd}^{(1)}\int U_{dd_{1}}^{(3D)}(\textbf{r}-\textbf{r}')|\Psi_1(\textbf{r}',t)|^{2}d\textbf{r}'
\\
&+g_{12}^{(3D)}|\Psi_2(\textbf{r},t)|^{2}+g_{dd}^{(12)}\int U_{dd_{12}}^{(3D)}(\textbf{r}-\textbf{r}')|\Psi_2(\textbf{r}',t)|^{2}d\textbf{r}')
\big\}
\Psi_1(\textbf{r},t),
\ea
\label{eq6}
\ee
\be
\ba{rl}
i\frac{\partial \Psi_2(\textbf{r},t)}{\partial t}=&
\big\{
-\frac{m_{12}}{2}\nabla^2+\frac{m_{\omega}}{2}(\lambda_2^2 x^2+\rho^{2})+g_2^{(3D)}\lvert\Psi_2(\textbf{r},t)|^{2}+g_{dd}^{(2)}\int U_{dd_{2}}^{(3D)}(\textbf{r}-\textbf{r}')|\Psi_2(\textbf{r}',t)|^{2}d\textbf{r}'\\
&+g_{21}^{(3D)}|\Psi_1(\textbf{r},t)|^{2}+g_{dd}^{(21)}\int U_{dd_{12}}^{(3D)}(\textbf{r}-\textbf{r}')|\Psi_1(\textbf{r}',t)|^{2}d\textbf{r}')
\big\}
\Psi_2(\textbf{r},t),
\ea
\label{eq7}
\ee
where $m_{12}=\frac{m_{1}}{m_{2}}$, $m_{\omega}=\frac{\omega_{2\bot}^{2}}{m_{12}\omega_{1\bot}^2}$, $g^{(3D)}_{1}=4\pi a_{1}N_{1}$, $g^{(3D)}_{2}=4\pi a_{2}N_{2}m_{12}$, $g_{dd}^{(1)}=3N_{1}a_{dd}^{(1)}$, $g_{dd}^{(2)}= 3N_{2}a_{dd}^{(2)} m_{12}$, $g^{(3D)}_{12}=\frac{2\pi a_{12}m_{1}N_2}{m_{r}}$, $g^{(3D)}_{21}=\frac{2\pi a_{12}m_{1}N_1}{m_{r}}$,  
$g_{dd}^{(12)}=3N_{2}a_{dd}^{(12)}$, $g_{dd}^{(21)}=3N_{1}a_{dd}^{(12)}$. The atom numbers $N_1$ and $N_2$ have been introduced in order for the wave functions to be normalized to unity.

In the system (\ref{eq6}-\ref{eq7}), lengths are expressed in units of the oscillator length $l_1=\sqrt{\frac{\hbar}{m_1\omega_{1\bot}}}$ and densities in units of $l_{1}^{-3}$. We also express energy and time in units of
$\hbar\omega_{1\bot}$ and $\omega_{1\bot}$ respectively. For the sake of clarity, we have also introduced length scales $a_{dd}^{(i)}$ and $a_{dd}^{(12)}$ corresponding to the DDI defined by 
$\frac{\mu_0\mu_i^2}{4\pi}=\frac{3\hbar^2}{m_i}a_{dd}^{(i)}$ and $\frac{\mu_0\mu_1\mu_2}{4\pi}=\frac{3\hbar^2}{m_1}a_{dd}^{(12)}$.

For a cigar-shaped trap, the transverse degrees of freedom are frozen. The kinematics of the system can be 
considered as quasi-one dimensional. In this situation, a good approximation is to split the individual wave 
functions into products of ground state functions of the harmonic oscillator (in the $\rho$ direction) and 
functions of $x,t$ alone. We may therefore take
\be
\Psi_i(\textbf{r},t)=\frac{1}{\sqrt{\pi l_{i}^{2}}}\exp(-\rho^{2}/2l_{i}^2)\psi_i(x,t).
\label{eq8}
\ee
Inserting the ansatz (\ref{eq8}) into the coupled GPE (\ref{eq6}-\ref{eq7}), multiplying both sides by $\exp(-\rho^2/2l_i^2)$ and integrating over $\rho$, we get the final quasi-1D coupled GPE:
\be 
\ba{rl}
i \frac{\partial \psi_1(x,t)}{\partial t}=&\big[
-\demi\nabla_x^2+\demi\lambda_1^2 x^2+\frac{g^{(3D)}_{1}}{2\pi l_1^2} |\psi_1 (x,t)|^2+g_{dd}^{(1)}\int U_{dd_1}^{(1D)}(x-x')|\psi_1 (x',t)|^2 dx'\\
& +\frac{g^{(3D)}_{12}}{\pi (l_1^2+l_2^2)}|\psi_2 (x,t)|^2+g_{dd}^{(12)}\int U_{dd_{12}}^{(1D)}(x-x')|\psi_2 (x',t)|^2 dx'
\big]
\psi_1 (x,t),
\ea
\label{eq18}
\ee

\be
\ba{rl}
i\frac{\partial \psi_2(x,t)}{\partial t}=&\big[
-\frac{m_{12}}{2}\nabla^2_x +\frac{m_{\omega}}{2}\lambda_2 x^2 +\frac{g^{(3D)}_{2}}{2\pi l_{2}^{2}} |\psi_2 (x,t)|^2 +g_{dd}^{(2)}\int U_{dd_{2}}^{(1D)}(x-x')|\psi_2 (x',t)|^2 dx'\\
& +\frac{g^{(3D)}_{21}}{\pi (l_1^2+l_2^2)}|\psi_1 (x,t)|^2 + g_{dd}^{(21)}\int U_{dd_{12}}^{(1D)}(x-x')|\psi_1 (x',t)|^2 dx'
\big]
\psi_2 (x,t).
\ea
\label{eq19}
\ee
In these equations, $U_{dd_{i}}^{(1D)}$ are the quasi-1D intra-species DDI given by \cite{Santos, baranov, 
Deuretzbacher, yukalov,  Muruganandam,  kanjilal}
\be
U_{dd_i}^{(1D)}(v_i)=\frac{1+3\cos2\alpha_i}{4l_i^{3}}\left(\frac{4}{3}\delta (v_i)+2\sqrt{v_i}-\sqrt{\pi}(1+2v_i)e^{v_i}\text{erfc}(\sqrt{v_i})\right),
\label{eq17}
\ee
where $v_i=(x-x')^2/l_i^2$, $\alpha_i$ are the angles formed by the magnetic moments and the $x$ axis and erfc 
is the complementary error function. Moreover, $U_{dd_{12}}^{(1D)}$ is the quasi-1D inter-species DDI which can 
be shown to write as (see appendix for more details)
\be
U_{dd_{12}}^{(1D)}(v_r)=\frac{\cos(\alpha_1-\alpha_2)+3\cos(\alpha_1+\alpha_2)}{2(l_1^2+l_2^2)^{3/2}}
\left(\frac{4}{3}\delta (v_r)+2\sqrt{v_r}-\sqrt{\pi}(1+2v_r)e^{v_r}\text{erfc}(\sqrt{v_r})\right),
\label{eq16}
\ee
where $v_r=(x-x')^2/(l_1^2+l_2^2)$. One notices that the expressions (\ref{eq17}) and (\ref{eq16}) are symmetric around $\alpha_i =\pi /2$ and become equivalent when the tilting angles are equal. Furthermore, the three dipolar potentials are attractive when $\alpha_i\le 0.5\arccos{(-1/3)}$ and repulsive otherwise. This will have great implications in the following.

\section{Numerical results}

In order to solve the coupled GPE (\ref{eq18}-\ref{eq19}), we use imaginary-time propagation with Crank Nicolson 
method as depicted in \cite{adhikari2, Adhikari2014, Muruganandam, sadhan}. This provides the ground state solutions 
of the mixture. The DDI are evaluated by means of a fast Fourier transform\cite{goral}. 

One may notice on (\ref{eq18}-\ref{eq19}) that the static properties depend on a large number of control parameters, including the number of particles in each species, the strengths of the two types of interactions (contact and DDI), 
the trap geometry as well as the orientations of the dipoles. Since the former parameters have been considered elsewhere\cite{Adhikari, kumar, Adhikari2014}, we will focus in the following on the effects of the tilting angles. 
Moreover, in order to be as close as possible to the experimental situations, and since it is quite difficult if not impossible to experimentally vary the angles independently\cite{gora}, we will for simplicity take  
$\alpha_1=\alpha_2=\alpha$ which will then be varied from $0$ up to $\pi/2$.

In the binary mixture, $^{168}$Er is the species labeled by 1, with $\mu_1= 7\mu_B$ ($\mu_B$ the Bohr magneton), $a_{dd}^{(1)}=66a_0$ ($a_0$ the Bohr radius). For $^{164}$Dy, we take $\mu_2= 10\mu_B$ and $a_{dd}^{(2)}=131a_0$ 
and this gives $a_{dd}^{(12)}=25a_0$ \cite{Mingwu}. In order to apprehend better the effects of the dipolar 
interactions on the miscibility-immiscibility transition, we deliberately choose repulsive and equal intra-species 
contact interactions: $a_1^{(3D)}=a_2^{(3D)}=200a_0$. The intra-species contact interactions are also repulsive $a_{12}^{(3D)}=140a_0$. Have we chosen $a_{12}^{(3D)}<0$, the net total interaction for $\alpha=0$ would have been attractive and consequently we will only get mixed configurations.
Moreover, the traps have equal parameters: $\omega_{1\bot}=\omega_{2\bot}=480\pi$, $\lambda_1=\lambda_2=0.2$.

We begin by fixing $N_1=2000$ and plot the densities (measured in units of $1/l_1$) and the energies of 
the two species $|\psi_i(x)|^2$ for $N_2=500$ as the tilting angle is varied from $\alpha=0$ where the DDI 
are attractive, up to $\alpha=\pi /2$ where they become repulsive. The DDI change their signs for the "magic" 
angle $\alpha^*=\demi\arccos{(-1/3)}$ which is around $\pi/3$.

Figure 1 depicts the densities of the two species for various titling angles. For $\alpha=0$, although the DDI 
are attractive, they are not strong enough to balance the repulsive contact forces. The net result is a lower 
energy for the Dy as shown in figure 2. The Er density partially splits into two symmetrical parts around the 
center of the trap, which leads to a quasi-demixed configuration.
However, as soon as the tilting angle approaches and then goes beyond $\alpha=\pi/6$, this spliting becomes 
less acute witnessing a tendency to mixing. 
Indeed, when $\alpha$ goes beyond the magic angle $\alpha^*$, the DDI vanish and then turn repulsive, 
being maximally repulsive for $\alpha=\pi/2$. This leads to an increasing positive energy for both species 
and therefore to an almost total mixing due to the spreading and flatening of the Dy. 


\begin{figure}[h!]
   \centering
   \includegraphics[scale=0.6]{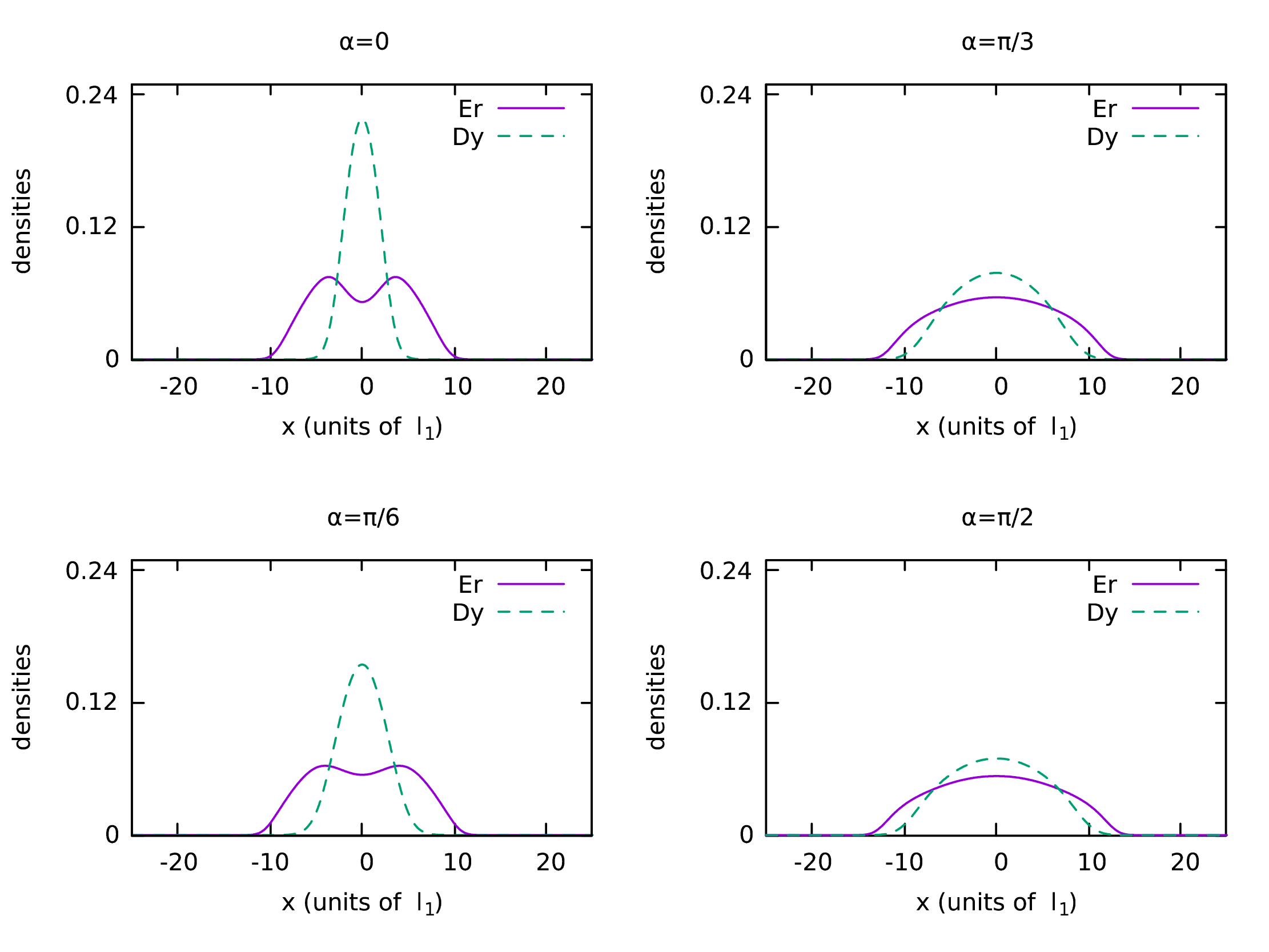}
   \caption{Density profiles for $^{168}$Er ($N_1=2000$, purple-continuous) and $^{164}$Dy ($N_2=500$, green-dashed).}
   \includegraphics[scale=0.5]{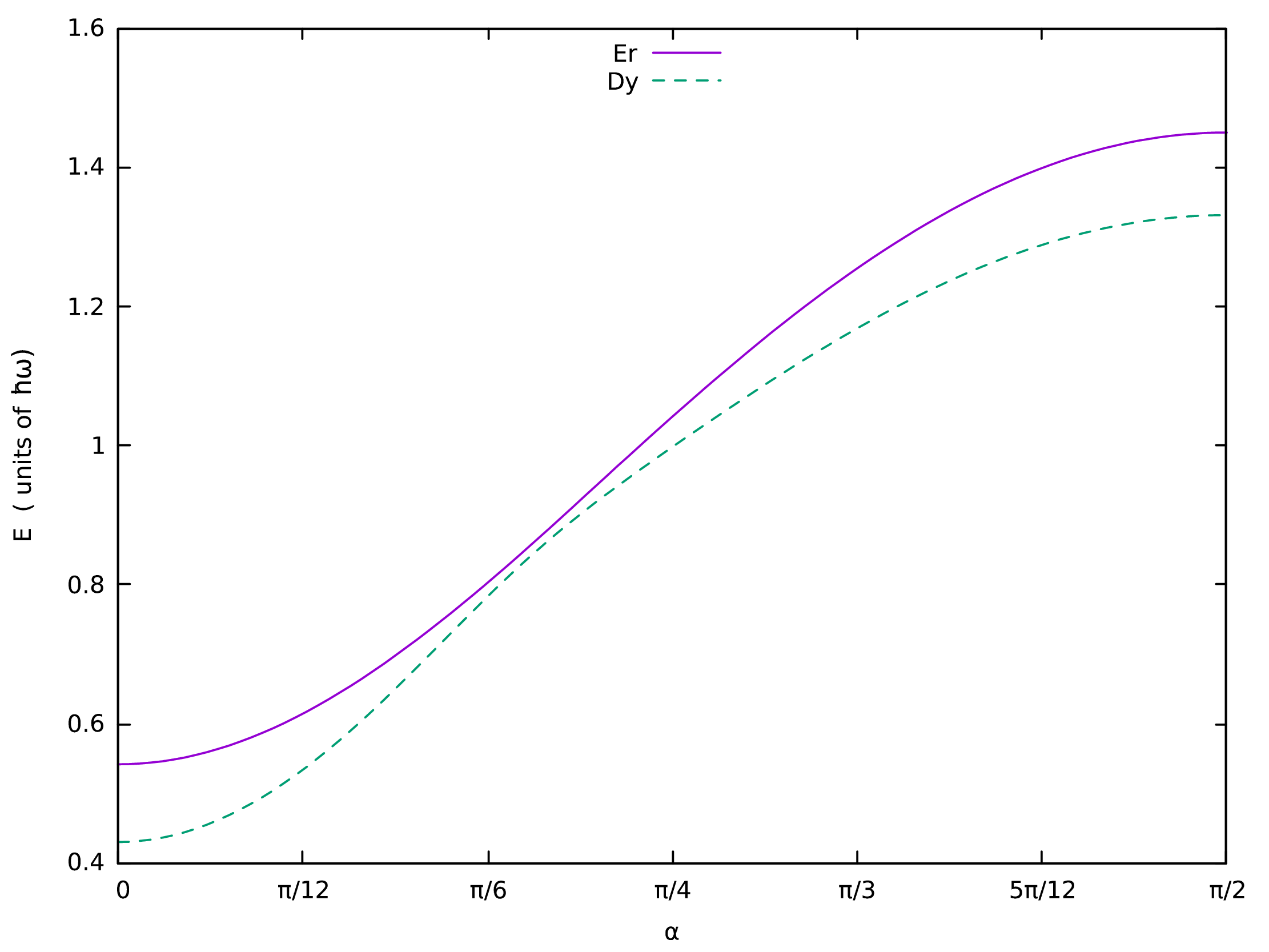}%
   \caption{Individual energies for $^{168}$Er ($N_1=2000$, purple-continuous) and $^{164}$Dy ($N_2=500$, green-dashed).}
\end{figure}

In order to see how the mixing-demixing transition is affected by the number of atoms, the calculations are repeated for 
$N_2=2000$ (figures 3 and 4) and $N_2=5000$ (figures 5 and 6). In these situations, the intra and inter-species 
interactions become comparable in absolute values. In the figures 3 and 5, a partial demixing is occurring at $\alpha=0$ 
and $\pi/6$, but now the Er density falls to zero around the center which means that the Dy atoms are occupying the center of the trap as their energy is lower, and are surrounded by the Er atoms which are then expelled at the peripheries. 
This partial demixing is more acute for growing Dy atom numbers, since, unlike the figure 3, in the case $N_2=5000$, 
there is a clear phase separation as the two peaks of the Er density become more distant and the region where there is 
no Er atoms becomes wider.

For repulsive DDI, ($\alpha=\pi/3$, $\pi/2$), as the energies of the two species become closer (see figures 4 and 6), 
one observes a tendency to mixing leading to an almost total mixing for $\alpha=\pi /2$. This argument is clearly the 
most physical one. The argument of masses, reported in \cite{kumar} can no longer be viable since it is most natural 
to compare the $a_{dd}$'s of the two species.



\begin{figure}[h!]
   \centering
  \includegraphics[scale=0.6]{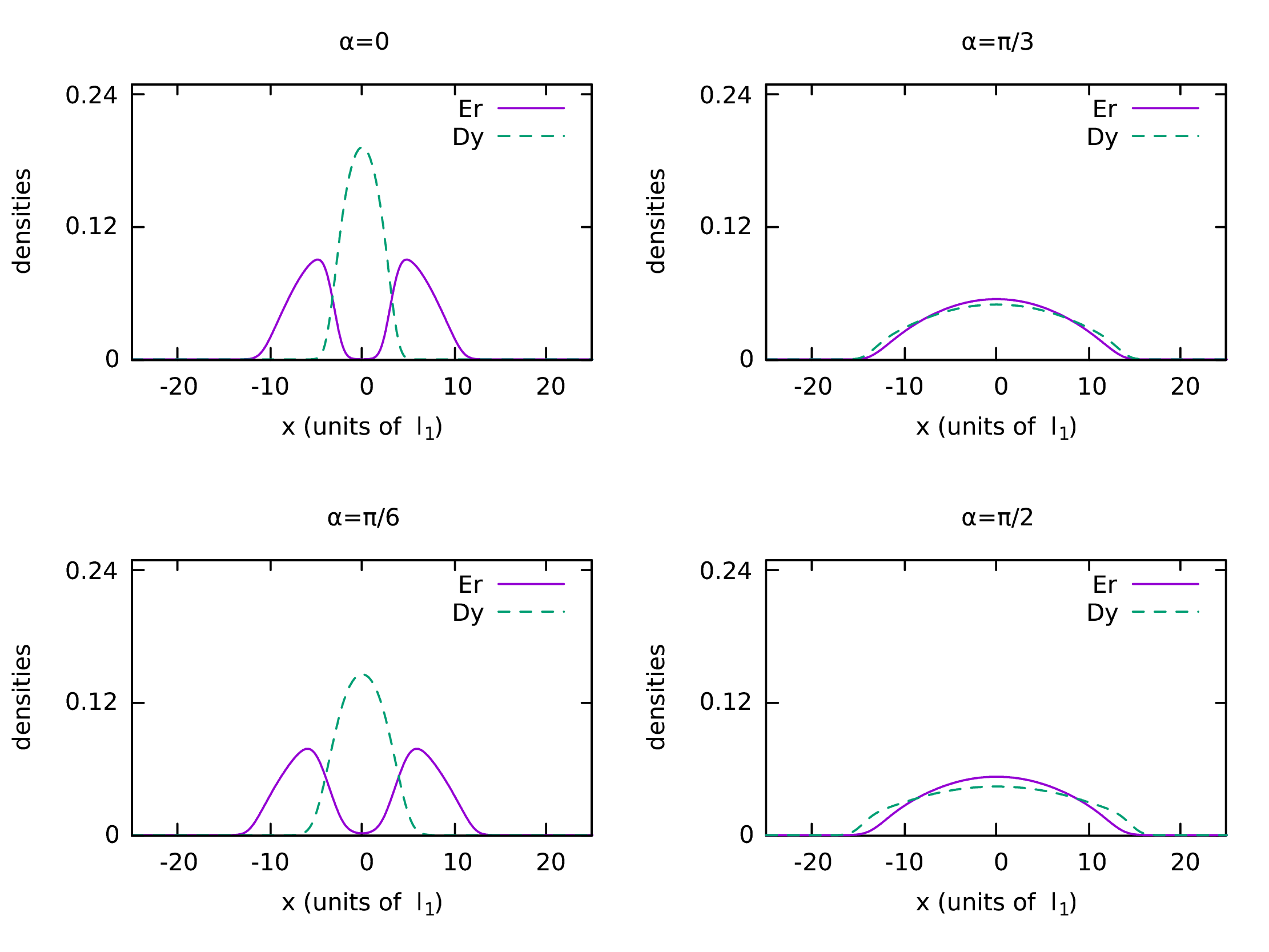}
   \hspace{1cm}%
   \caption{Density profiles for $^{168}$Er ($N_1=2000$, purple-continuous) and $^{164}$Dy ($N_2=2000$, green-dashed).}
   \includegraphics[scale=0.5]{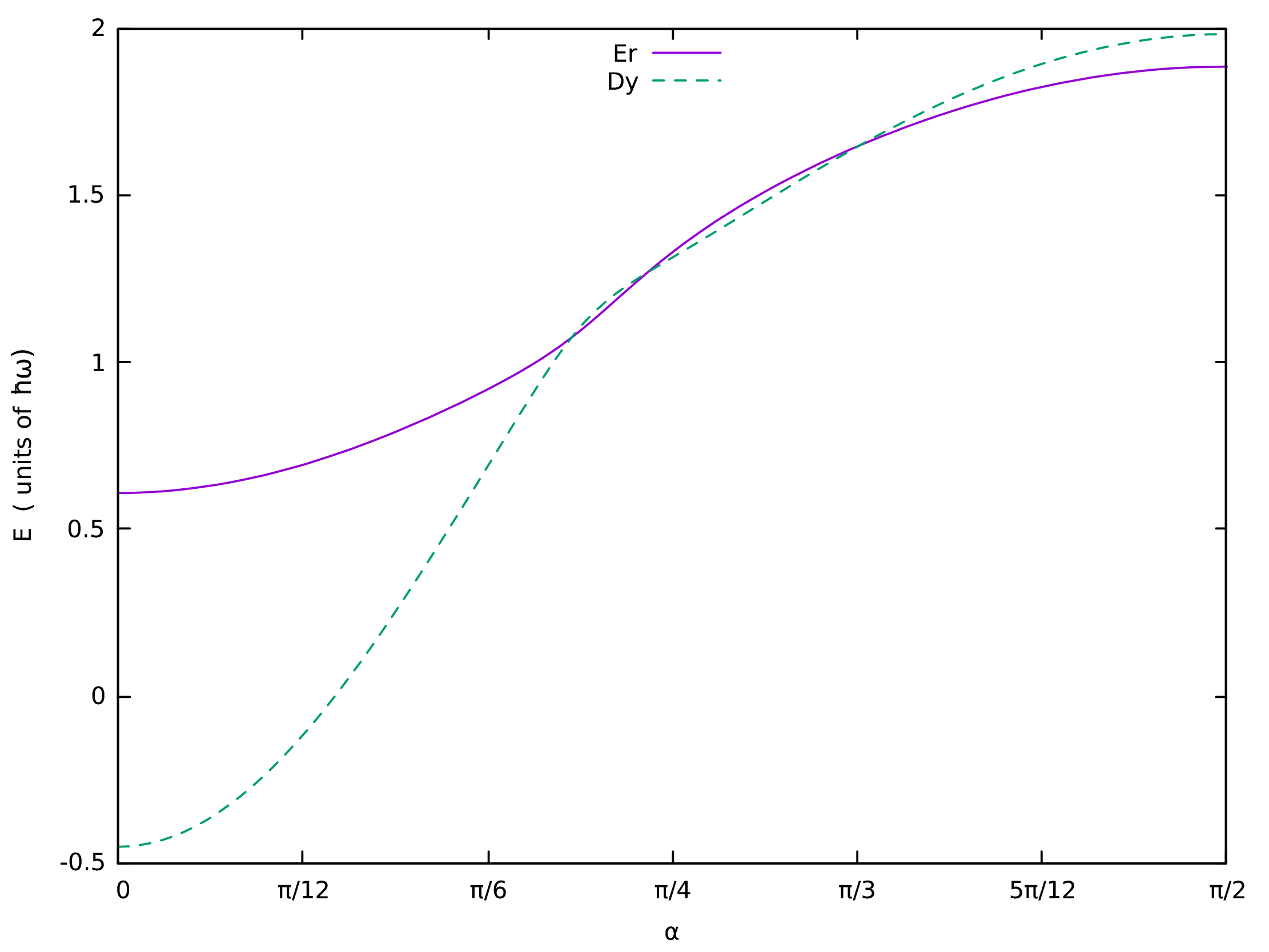}
   \hspace{1cm}%
   \caption{Individual energies for $^{168}$Er ($N_1=2000$, purple-continuous) and $^{164}$Dy ($N_2=2000$, green-dashed).}
\end{figure}

Furthermore, one observes on figure 6 a remarkable variation of the Dy energy around $\alpha=\pi/4$. 
This brutal variation is not evident on the density profiles shown in figure 5. For these orientations 
of the magnetic moments, the attractive and repulsive parts of the DDI are equal. Since there is no net 
attractive component, the interaction becomes suddenly repulsive due to the high number of Dy atoms.

\begin{figure}[h!]
   \centering
   \includegraphics[scale=0.6]{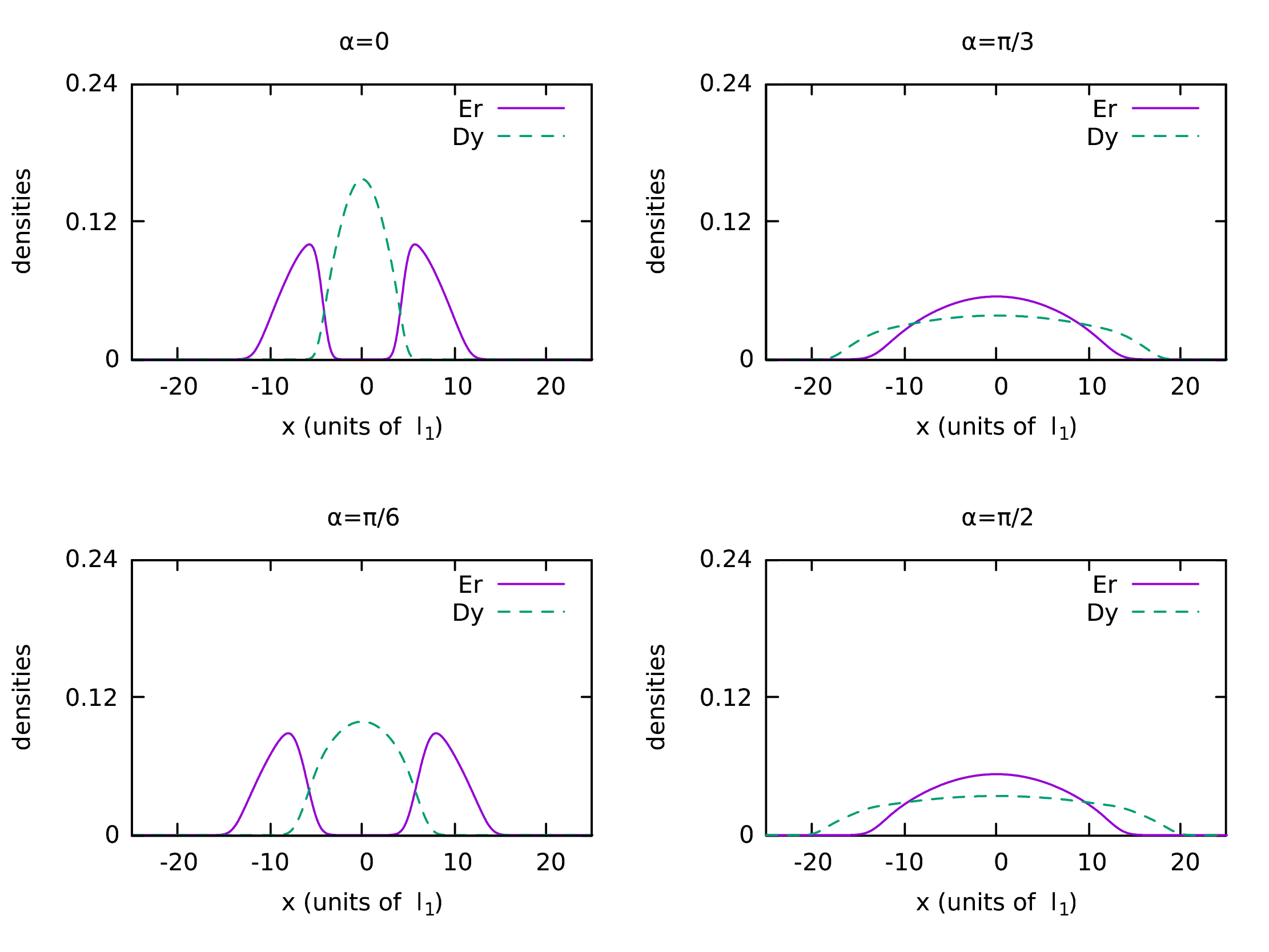}
   \hspace{1cm}%
   \caption{Density profiles for $^{168}$Er ($N_1=2000$, purple-continuous) and $^{164}$Dy ($N_2=5000$, green-dashed).}
   \includegraphics[scale=0.5]{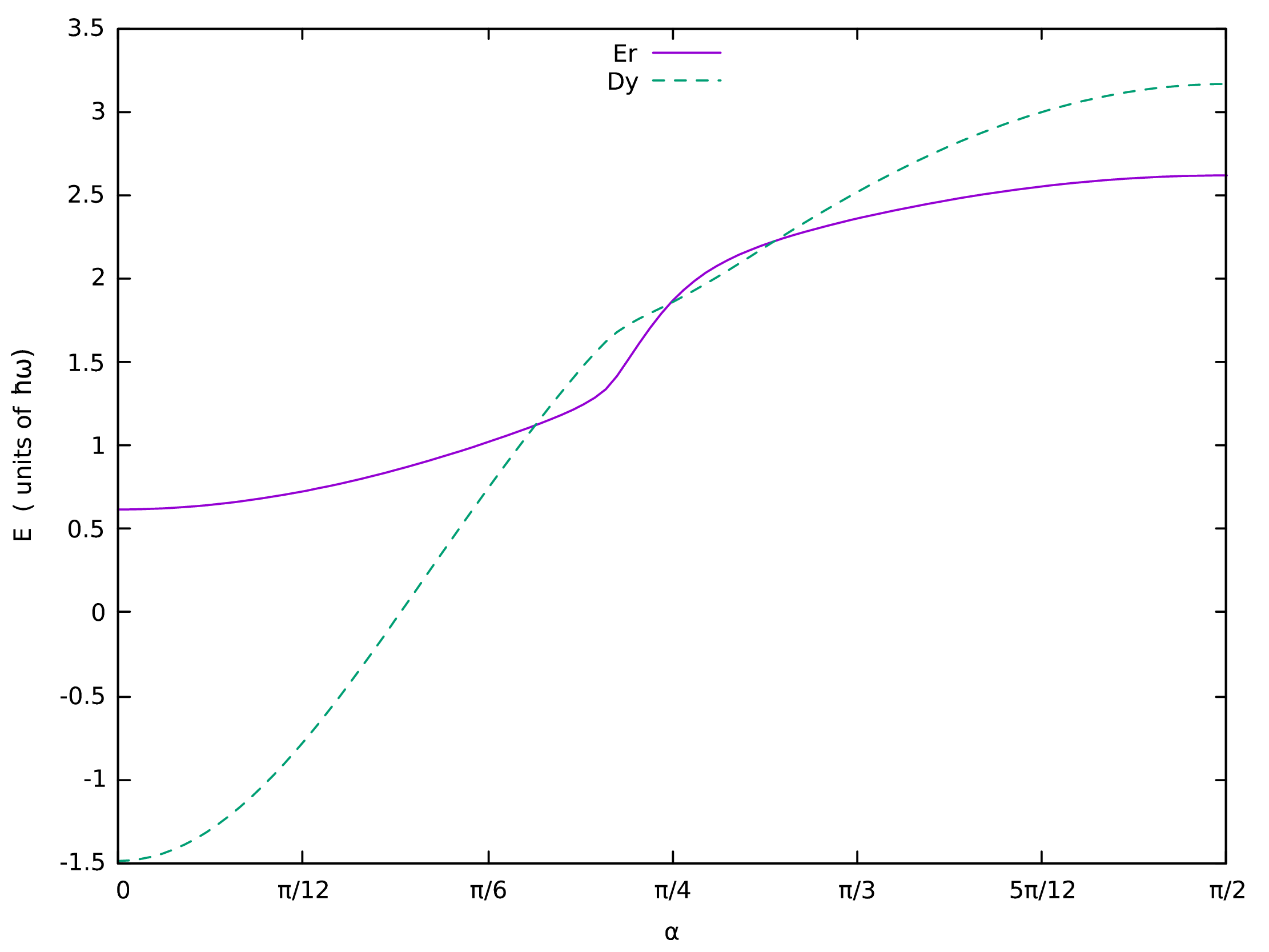}
   \hspace{1cm}%
   \caption{Individual energies for $^{168}$Er ($N_1=2000$, purple-continuous) and $^{164}$Dy ($N_2=5000$, green-dashed).}
\end{figure}


This behavior may be well illustrated by the overlap (or miscibility) parameter 
$\eta=\int_x\sqrt{|\psi_1(x)|^2|\psi_2(x)|^2}$ as defined in \cite{kumar}. 
This quantity is bestly suited to the inhomogenous case instead of the parameter $\Delta$ 
(see \cite{kumar} Eq.9).

The figure 7 depicts $\eta$ as a function of Dy atom number for different tilting angles $\alpha$. 
This quantity clearly varies from very small values for $\alpha=0$, where the DDI are maximally attractive 
leading to a partially demixed system, up to almost 1 (for $\alpha=\pi /2$), where the DDI are maximally 
repulsive and the system is totally mixed.

\begin{figure}[h!]
  \centering
   \includegraphics[scale=0.5]{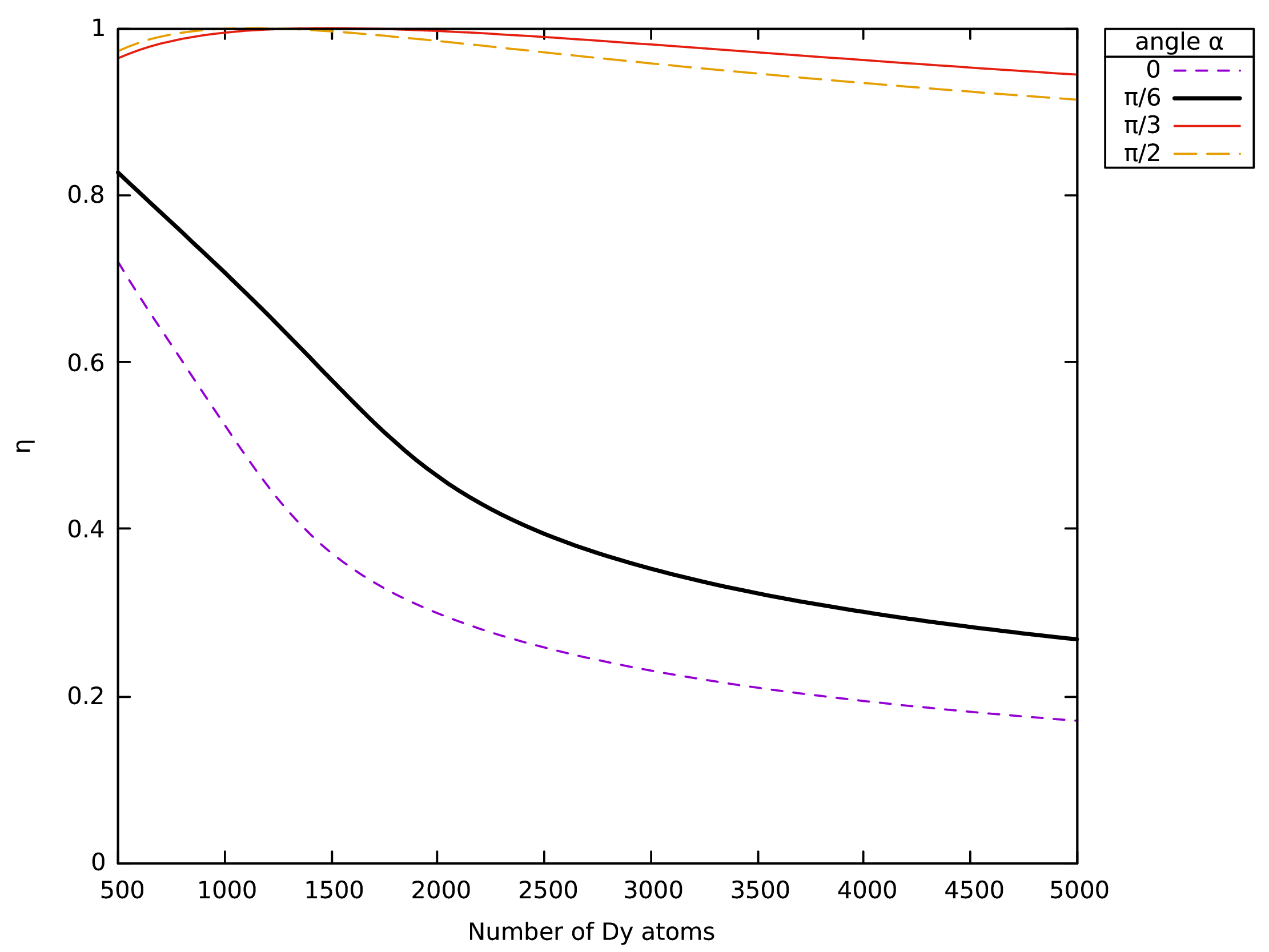}
  \hspace{1cm}%
   \caption{Overlap parameter for $N_1=2000$ atoms of $^{168}$Er versus the $^{164}$Dy atom number $N_2$ for various 
   titling angles.}
\end{figure}

In the figure 8, we represent $\eta$ as a function of $\alpha$ for different Dy atom numbers. On this figure, 
we observe a strong partial demixing for small tilting angles, which we are tempted to call an immiscible 
configuration. For growing angles, the overlap nearly reaches its maximum value (depending on $N_2$) in the strongly repulsive DDI case ($\alpha=\pi /2$).

A noticeable feature is the fixed point (around ($\alpha=\pi /5$) which indicates that the overlap crosses a constant 
value whatever the number of Dy atoms. This result comes from the very definition of $\eta$ since $\psi_2$ is normalized 
to unity.

\begin{figure}[h!]
  \centering
   \includegraphics[scale=0.5]{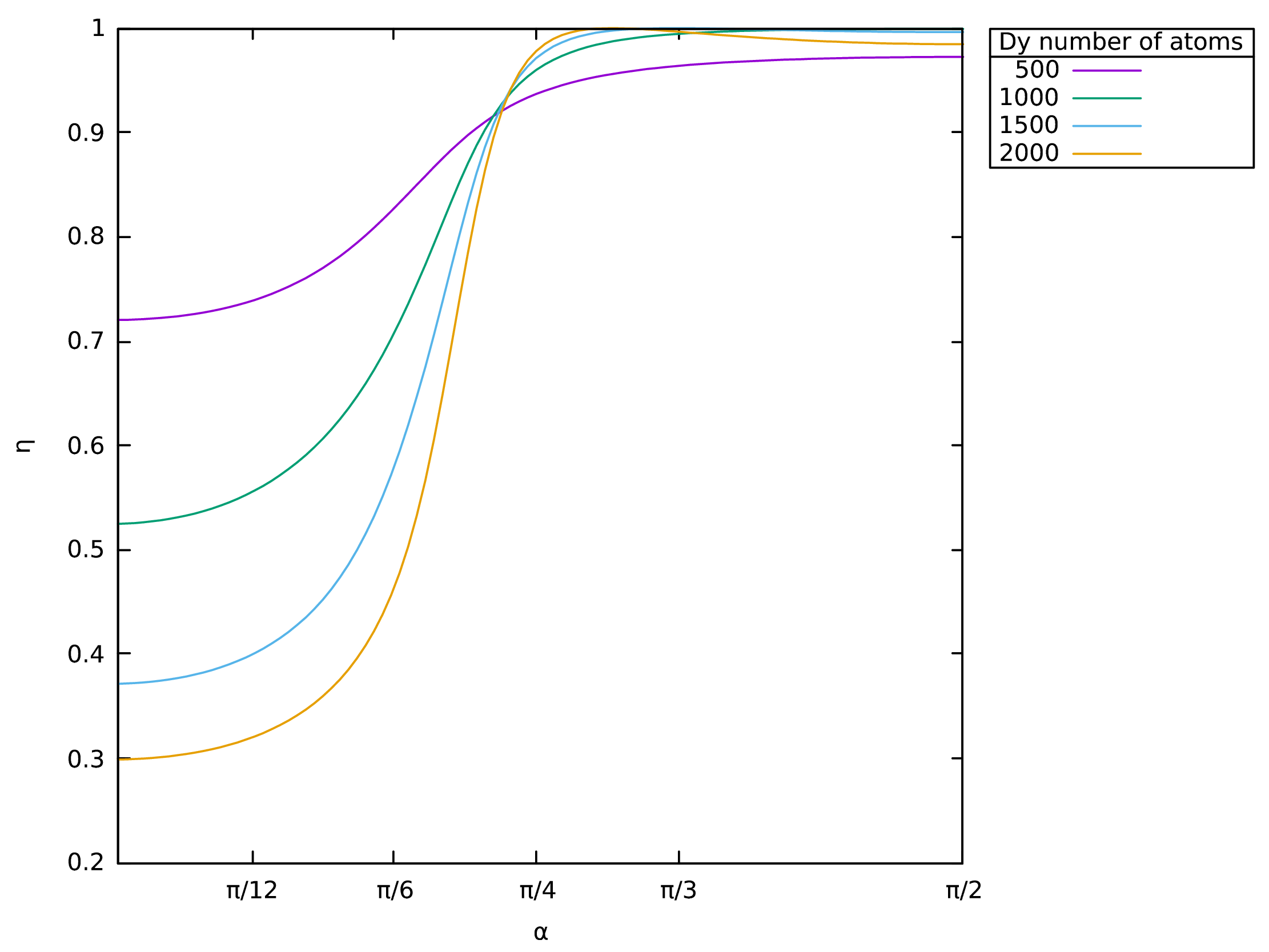}
  \hspace{1cm}%
   \caption{Overlap parameter for $N_1=2000$ atoms of $^{168}$Er versus the tilting angle for various $^{164}$Dy atom 
   number $N_2$.}
\end{figure}
\newpage

\setcounter{equation}{0}
\section{Concluding remarks}
\renewcommand{\theequation}{\arabic{equation}}

In the present work, we have studied a zero temperature dipolar bose mixture, namely $^{168}$Er-$^{164}$Dy,
trapped in quasi-1D geometry in the mean field approximation. We have been particularly concerned with the 
effect of the orientation of the magnetic moments on the transition from miscibility to immiscibility.

Upon solving numerically the coupled Gross-Pitaevskii equations, we have explored various situations where one 
changes the tilting angles as well as the number of atoms. 

The outcomes of our calculations show that a transition from demixed to mixed configuration can be driven 
by changing the orientations of the dipoles and therefore changing the DDI from attractive to repulsive.
This behavior is particularly marked for growing atom numbers, where we observe that the less energetic 
species occupy the center of the trap while expelling the more energetic one at the peripheries. 
Indeed, the density of the latter shows two symmetric peaks around the center separated by a void region 
which becomes wider as the number of atoms of the former species grows up.

Moreover, upon examining the overlap of the macroscopic wave functions of the two species, we notice a clear 
tendency to maximum mixing due to the repulsive DDI. The demixing is however partial since within the parameter 
space that we have considered, the immiscibility conditions are not fully reached.

Our results are not only consistents with previous works\cite{kumar, Adhikari2014} but also extend their claims.
Indeed, we found that the mixture can be made {\it completely} miscible (the overlap almost reaching its maximum 
value) by just changing the orientations of the magnetic moments of the two species while keeping fixed all the 
other parameters.
Based on energy arguments, we conclude that it is the less energetic species, with an attractive dipolar 
component, that will occupy the center of the trap. The argument of masses invoked in \cite{kumar} is no 
longer viable.
Moreover, in the quasi-1D geometry, the stability window, in terms of the fraction of atom numbers or 
trap anisotropy, is found to be wider accomodating for higher numbers of atoms in each species and 
therefore almost reaching the Thomas-Fermi regime at very low aspect ratios.
This result can be helpful for future experimental setups.

As an interesting perspective, it would be quite natural to examine the excitation spectrum of this 
quasi-1D mixture. Indeed, the emergence of an anisotropic roton mode, with very special properties 
as a function of the polarization angle has been predicted in \cite{wilson} for a quasi-2D traps. 
The question is to what extent such a mode may survive in reduced geometries. This and other related 
questions will be addressed in a forthcoming paper.

We would like to thank S. Adhikari for valuable comments about the manuscript. 
G. Shlyapnikov and D. Petrov are aknoweledged for fruitful discussions and kind hospitality at 
LPTMS-Paris.

\setcounter{equation}{0}
\renewcommand{\theequation}{\arabic{equation}}
{\Large {\bf Appendix}}

In this appendix, we provide the details for the computation of the quasi-1D effective potential (\ref{eq16}) 
which appears in the Eqs. (\ref{eq18}-\ref{eq19}). It is given by the integral expression:
\be
U^{(1D)}_{dd_{12}}(x_1-x_2)=\frac{1}{\pi^{2}l_{1}^2 l_{2}^2}\int d^2\rho_1 d^2\rho_2 dx_2 U_{dd_{12}}^{(3D)}(\textbf{r}_1-\textbf{r}_2)\exp(-\rho_1^2 /l_1^2)\exp(-\rho_2^2 /l_2^2),
\label{eq20}
\ee
where, up to a factor $\mu_0\mu_1\mu_2/4\pi$, the 3D potential is given by \cite{Lahaye}
\be
U_{dd_{12}}^{(3D)}(\textbf{r})=\frac{(\textbf{e}_{1}.\textbf{e}_{2})r^2 -3(\textbf{e}_{1}.\textbf{r})(\textbf{e}_{2}.\textbf{r})}{r^{5}}.
\label{eq26}
\ee
$\textbf{e}_{1}$ and $\textbf{e}_{2}$ are the unit vectors along the dipole moments directions with $\frac{\textbf{e}_{i}.\textbf{r}}{r}=\cos\theta_{i}$ and $\textbf{e}_{1}.\textbf{e}_{2}=\cos(\theta_{1}-\theta_{2})$ which leads to (\ref{eq331}). 

In order to compute the integrals appearing in (\ref{eq20}), we introduce the relative ($y$, $z$) and the center of mass (CM) cartesian coordinates ($Y$, $Z$) in the $y-z$ plane:
\be
\ba{rl}
y_{1,2}=&Y\pm\frac{m_{2,1}}{m_1+m_2}y\\
z_{1,2}=&Z\pm\frac{m_{2,1}}{m_1+m_2}z 
\ea .
\label{eq21}
\ee
It is now straightforward to notice that the CM coordinates can be integrated out to yield an expression depending 
solely on the relative coordinates. Noting $x=x_1-x_2$, we get
\be
U^{(1D)}_{dd_{12}}(x)={\ds 1\over\ds \pi(l_1^2+l_2^2)}\int dydz\, U_{dd_{12}}^{(3D)}(x, y, z)
\exp\left(-{\ds y^2+z^2\over\ds l_1^2+l_2^2}\right) .
\label{eq24}
\ee

Now assuming that the dipoles are in the $x-z$ plane, $\textbf{e}_i=(\cos\alpha_i,0,\sin\alpha_i)$, we may change back to polar coordinates ($y=\rho\cos\phi$, $z=\rho\sin\phi$). One obtains the simple relations
\be
\ba{rl}
\cos(\theta_1-\theta_2)=&\cos(\alpha_1-\alpha_2),\\
\cos\theta_i=&{\ds x\cos\alpha_i+\rho\sin\phi\sin\alpha_i\over\ds(x^2+\rho^2)^{1/2}},
\ea
\label{eq27}
\ee
which allow us to write (\ref{eq24}) in the simpler form 
\be
U^{(1D)}_{dd_{12}}(x)={\ds\cos(\alpha_1-\alpha_2)+3\cos(\alpha_1+\alpha_2)\over\ds 2(l_1^2+l_2^2)}\int_0^{\infty}d\rho \,\rho{\ds \rho^2-2x^2 \over\ds (x^2+\rho^2)^{5/2}}\exp\left(-{\ds\rho^2\over\ds l_1^2+l_2^2}\right) ,
\label{eq28}
\ee
and directly yields the result (\ref{eq16}).

\end{document}